\documentclass[aps,prl,twocolumn,superscriptaddress,showpacs]{revtex4}

\usepackage[dvips]{graphicx}
\usepackage{amsmath}
\usepackage{amssymb}
\usepackage{bm}

\begin{document}
\title{The Stability of the Replica Symmetric State in Finite Dimensional Spin
 Glasses }

\author{M. A. Moore}
\affiliation{School of Physics and Astronomy,
University of Manchester, Manchester M13 9PL, U.\ K.}

\date{\today}

\begin{abstract}
According to the droplet  picture of spin glasses, the low-temperature
phase of  spin glasses should be replica  symmetric. However, analysis
of the stability of this state suggested that it was unstable and this
instability  lends support  to  the Parisi  replica symmetry  breaking
picture  of spin  glasses.  The  finite-size scaling functions  in the
critical region of spin glasses below $T_c$ in dimensions greater than
6 can be determined  and for them the replica symmetric solution
is unstable  order by order in perturbation  theory.  Nevertheless the
exact solution can be shown  to be replica-symmetric.  It is suggested
that a similar  mechanism might apply in the  low-temperature phase of
spin glasses in less than  six dimensions, but that a replica symmetry
broken state might exist in more than six dimensions.

\end{abstract}



\maketitle

An important  development in statistical physics  was Parisi's replica
 symmetry breaking (RSB)  solution of the Sherrington-Kirkpatrick (SK)
 model of spin glasses \cite{P}.  The  SK model is the model for which
 mean-field theory  is exact, by virtue  of the infinite  range of the
 interactions between the spins.  The Parisi solution is characterised
 by  a large  number of  pure  states, organised  into an  ultrametric
 hierarchy \cite{BP}. While  there is no debate as  to whether the RSB
 solution is correct for the SK model, almost from the beginning there
 has  been a  controversy  as to  whether  it extends  to real  finite
 dimensional  spin glasses.   To extend  it to  finite  dimensions one
 constructs  the loop (perturbative)  expansion around  the mean-field
 solution.  At least in high dimensions this perturbative expansion is
 well-defined i.e.   all the integrals are finite.   This programme is
 entirely  conventional  and  most   of  our  understanding  of  phase
 transitions  and the  low-temperature phase  across  condensed matter
 physics  has  followed  this   route  (mean-field  theory,  then  the
 fluctuations around the mean-field). How  then could it fail for spin
 glasses? This is the question which we will attempt to answer in this
 paper.

The chief rival to the  Parisi replica symmetry breaking (RSB) picture
is the droplet picture \cite{droplet}.  On this picture there are just
two  pure states  (the  analogue of  the  up and  down  states of  the
ferromagnet),   and   so   the   low-temperature  phase   is   replica
symmetric. The properties of the  spin glass phase are governed by its
excitations; the free-energy cost of flipping the spins in a region of
linear extent $L$  is supposed to be $L^{\theta}$,  where the exponent
$\theta \sim 0.2$ in three dimensions, according to numerical studies.
The droplet has fractal area $L^{d_s}$ where $d-1\le d_s\le d$
and  $d$ is  the  dimensionality of  the  system.  The  droplet has  a
fractal  surface as  it wanders  to  pass the  domain wall  separating
reversed spins  through as  many weak bonds  as possible, in  order to
minimize its overall energy.  Thus the droplet picture focusses on the
low-cost excitations, which are  determined by unusual correlations of
the bonds, and attributes  the properties of the low-temperature phase
to these \lq Griffiths' singularities.

The droplets  are very intimately  connected to non-average  values of
the bonds between the spins  on nearby sites.  However, in the replica
trick the first  step is to average out the  bond disorder. Thus after
using the replica trick, features which can be associated with droplet
behavior  can   only  arise  from   a  treatment  which   goes  beyond
straightforward perturbation  theory.  In  the example which  we shall
discuss in detail  below, at each finite order  in the expansion about
the  replica symmetric  solution there  are instabilities  which might
lead one to believe that replica symmetry must be broken. However, the
exact  solution remains replica  symmetric.  It  is natural  to wonder
whether a  similar mechanism might  apply more generally and  we shall
give arguments as to why this might happen.

The replica field theory of spin  glasses, (see Ref.  \cite {KD} for a
review), starts from the Hamiltonian density
\begin{align}
&\mathcal{H}=\frac  14\sum_{\alpha,\beta}  (\nabla  q_{\alpha\beta})^2
           +\frac \tau 4\sum_{\alpha,\beta}q_{\alpha\beta}^2 -\frac w6
           \sum_{\alpha,\beta,\gamma}q_{\alpha\beta}q_{\beta\gamma}q_{\gamma\alpha}
\label{FT}\\
&-y\Big(\frac 1  {12} \sum_{\alpha,\beta}q_{\alpha\beta}^4 +\frac  1 8
   \sum_{\alpha,\beta,\gamma,\delta}q_{\alpha\beta}q_{\beta\gamma}
   q_{\gamma\delta}q_{\delta\alpha}-\frac              1             4
   \sum_{\alpha,\beta,\gamma}   q_{\alpha\beta}^2   q_{\alpha\gamma}^2
   \Big).\nonumber
\end{align}
The   field   components   $q_{\alpha\beta}$  ($\alpha\neq\beta$   and
$q_{\alpha\beta}=q_{\beta\alpha}$)  take  all  real  values,  and  the
indices such as $\alpha$ take the  values $1, 2, 3, \hdots, n$. In the
limit when  $n$ goes  to zero, it  is hoped  that such a  field theory
captures the physics of  finite dimensional spin glasses.  The quartic
terms are for $d<6$ irrelevant variables \cite{LC}, as they are in the
finite-size  critical  regime  in   any  dimension  so  they  will  be
dropped. The coefficient of the  quadratic term $\tau$ changes sign at
the   mean-field  transition  temperature   $T_{c0}$  so   $\tau  \sim
(T-T_{c0})/T_{c0}$.

We will start  by briefly reviewing the kind  of calculation which has
led  to the  view  that  the replica  symmetric  solution is  unstable
\cite{BM79}.   One begins  with  the mean-field  solution  which is  a
stationary point  of the Hamiltonian  density of Eq.   (\ref{FT}). The
standard replica symmetric solution (on dropping the quartic terms) is
the  spatially  uniform  $q_{\alpha\beta}({\bf x)}=Q$  independent  of
$\alpha$ and $\beta$, where $Q=\tau/(n-2)w$. Only the trivial solution
$Q=0$  exists on  the high-temperature  side of  the  transition where
$\tau  > 0$.   The stability  of this  replica symmetric  solution for
negative $\tau$  at and beyond  mean-field theory will occupy  most of
this paper.

The first step is to write
\begin{equation}
q_{\alpha\beta}=Q+R_{\alpha\beta}
\label{Fl}
\end{equation}
and substitute  into Eq. (\ref{FT}) (without the  quartic terms). Then
up to constants
\begin{align}
&\mathcal{H}\{R_{\alpha\beta}\}         =         \frac         12\tau
\sum_{\alpha<\beta}R^2_{\alpha\beta}+             \frac             12
\sum_{\alpha<\beta}(\nabla  R_{\alpha\beta})^2   \nonumber  \\  &  -wQ
\sum_{\alpha   <   \beta  <\gamma}   (R_{\alpha\beta}R_{\alpha\gamma}+
R_{\alpha\beta}R_{\beta\gamma}+R_{\alpha\gamma}R_{\beta\gamma})
\nonumber\\        &-w\sum_{\alpha       <        \beta       <\gamma}
R_{\alpha\beta}R_{\alpha\gamma}R_{\beta\gamma}.
\label{q2}
\end{align} 
The quadratic terms are not  diagonal. It is useful to first introduce
the  following   propagators  in  terms  of   the  Fourier  components
$R_{\alpha\beta}({\bf q})$
\begin{align}
&G_1(q)=\langle      R_{\alpha\beta}({\bf      q})R_{\alpha\beta}({\bf
  -q})\rangle,    \nonumber\\   &G_2(q)=\langle   R_{\alpha\beta}({\bf
  q})R_{\alpha\gamma}({\bf -q})\rangle,\hspace{0.5cm} \beta \ne \gamma
\nonumber\\            &G_3(q)=\langle            R_{\alpha\beta}({\bf
  q})R_{\gamma\delta}({\bf  -q})\rangle,  \hspace{0.5cm}  \alpha,\beta
\ne \gamma, \delta.
\label{defG}
\end{align}
Then,  following  Ref. (\cite{BM79})  the  quadratic  form is  readily
diagonalized in terms of three linear combinations of $G_1$, $G_2$ and
$G_3$:
\begin{align}
&G_B   \equiv   G_1+2(n-2)G_2+\frac  12(n-2)(n-3)G_3=(q^2+|\tau|)^{-1}
\nonumber\\     &G_A    \equiv    G_1+(n-4)G_2-(n-3)G_3=(q^2+2wQ)^{-1}
\nonumber\\ &G_R \equiv G_1-2G_2+G_3=(q^2+nwQ)^{-1}.
\label{rep}
\end{align}
All  three of these  propagators are  of the  form $(q^2+m^2_s)^{-1}$,
with the mass of the longitudinal mode given by $m_L^2=|\tau|$, of the
\lq anomalous'  mode by  $m_A^2=2wQ$ and of the replicon  mode by
$m_R^2=nwQ$. In  the limit of $n  \rightarrow 0$ the  breather and the
anomalous masses  become equal while  the replicon mass goes  to zero.
Stabilty  of course  requires that  all the  $m^2_s$  be non-negative.
Thus  at Gaussian order  the replica  symmetric solution  has marginal
stability.  (If we  had retained the quartic terms  in the Hamiltonian
density  the replicon  mode  would have  become  unstable at  Gaussian
order).   To see  the apparent  instability of  the  replica symmetric
state  it is  necessary to  go  to one  loop order  and calculate  the
self-energies of  the propagators.  The replicon self-energy
$\Sigma_R(q)$ is defined via
\begin{equation}
G_R=(q^2+nwQ-\Sigma_R(q))^{-1}.
\label{SED}
\end{equation}

To one-loop order the  calculation of $\Sigma_R(q)$ is straightforward
\cite{BM79}, \cite{TDP},
\begin{align}
&\Sigma_R(0)=\frac{n^4-8n^3+19n^2-4n-16}{(n-1)(n-2)^2}
I_{RR}\label{SE}\\
&+\frac{8(n-1)(n-4)}{n(n-2)^2}I_{RA}+\frac{8}{n(n-1)}I_{RL}
+\frac{(n-4)^2}{(n-2)^2}I_{AA},\nonumber
\end{align}
where for the wavevector sums we have introduced the notation
\begin{equation}
I_{ss'}=\frac    {w^2}{N}     \sum_{{\bf    q}}    \frac{1}{q^2+m^2_s}
\frac{1}{q^2+m^2_{s'}}
\label{bub}
\end{equation} 
and $s$ and  $s'$ correspond to one  of the subspaces R, A,  L. In the
limit  $n  \rightarrow  0$,   $\Sigma_R(0)$  can  be  approximated  as
\cite{BM79}
\begin{equation}
\Sigma_R(0)          \approx          \frac{4w^2|\tau|^2}{N}\sum_{{\bf
q}}\frac{1}{(q^2+nwQ)^2(q^2+|\tau|)^2}.
\label{SEA}
\end{equation}
In the large  $N$ limit the sum over the wavevectors  ${\bf q}$ in Eq.
 (\ref{SEA}) can be converted to  an integral. For $d>8$ the integrals
 will exist if cutoff at $q=\Lambda$, where $\Lambda \sim 1/a$ and $a$
 is the lattice spacing.   Then $\Sigma_R(0) \sim |\tau|^2$ on setting
 $n$ to zero.  For $4<d<8$,  in the same limit, $\Sigma_R(0)$ does not
 require an upper cutoff and $\Sigma_R(0) \sim |\tau|^{(d-4)/2}$. When
 $d<4$, the  integral is  dominated by its  small $q$ behavior  and is
 only finite if the replicon mass  is kept finite (i.e. by not setting
 $n$ to zero):
\begin{equation}
 \Sigma_R(0) \sim \frac{1}{|nwQ|^{(4-d)/2}}.
\label{ndep}
\end{equation}

For all  dimensions to this order $\Sigma_R(0)$  is positive, implying
that  $m_R^2$ is  negative  and  that at  one-loop  order the  replica
symmetric state  is unstable. However,  in an earlier paper  we argued
that such  a conclusion was  premature \cite{MB85}.  One  expects that
deep  within  the  ordered  phase, that  is,  when  $|\tau|\rightarrow
\infty$, loop  corrections should be small.  However,  as noted above,
below four  dimensions $\Sigma_R(0)$ actually becomes  infinite as $n$
goes to zero.  Worse divergencies exist in higher-loop corrections, as
diagrams exist which diverge whenever  $d<6$ and $n$ goes to zero.  We
suggested  therefore  that it  might  be  incorrect  to conclude  that
replica  symmetry had  to be  broken for  dimensions $d<6$,  since the
higher the  loop correction, the  more divergent the diagram  and that
one had  to go beyond order-by-order perturbation  theory in dimension
$d<6$ in  order to  discover whether the  replica symmetric  state was
really unstable \cite{MB85}.

Additional arguments were  also given in Ref. (\cite{MB85})  as to why
six might be a special dimension for the nature of the low-temperature
spin  glass  state.  If  it  is  indeed  replica symmetric  below  six
dimensions  then  there  should   be  no  Almeida-Thouless  (AT)  line
\cite{AT} as the AT line marks the onset of replica symmetry breaking.
This is consistent with  the renormalization group calculation of Bray
and Roberts \cite{BR}  who were unable to locate  a stable fixed point
in $6-\epsilon$  dimensions and as a consequence  speculated that this
could  be  due  to the  fact  that  there  was  no  AT line  in  these
dimensions.   ( Simulations  of  three dimensional  spin glasses  seem
consistent   with  this  conclusion   \cite{YK},  as   do  experiments
\cite{PN95}).

For dimensions $d>6$ the loop  expansion about the {\it RSB mean-field
state} is well-behaved \cite{KD} and one might therefore hope that the
Parisi RSB picture for these  dimensions is a valid description of the
spin  glass  state.  Below  six  dimensions,  the  loop expansion  has
divergences  which were  related  to the  divergences associated  with
non-mean-field  critical exponents  \cite{KD}. The  loop  expansion is
meant to be used well away from  the critical region, so it is hard to
understand  this identification.   It is  possible that  they indicate
instead that the RSB state is unstable in less than six dimensions.

Our  old argument,  that  calculations  to finite  order  in the  loop
expansion  might not  predict  the correct  stability  of the  replica
symmetric  state  when  coefficients  in  the  loop  expansion  become
infinite as  $n$ goes to  zero \cite{MB85}, met with  little interest,
perhaps  because no  concrete realization  could be  given. I  can now
supply an  example where,  order by order  in the loop  expansion, the
coefficients diverge  as $n$ goes  to zero, have signs  which indicate
that the replica symmetric state is apparently unstable, but which can
be  exactly  resummed to  a  solution  which  shows that  the  replica
symmetric state is stable, which  parallels what is being argued might
happen in spin glasses for $d<6$.

 This illustrative  calculation is of the low-temperature  form of the
finite size  critical scaling functions  for $d>6$, which can  be done
\lq  exactly'. It  is as  follows. One  supposes that  the  spin glass
system is of finite linear extent $L$ and that the system has periodic
boundary  conditions.  $N=L^d$.   Because  of the  periodicity of  the
system, the order parameter will be still uniform.  One can proceed to
construct  a  loop  expansion  as  for the  infinite  system,  and  in
expressions like  Eq.  (\ref{SE})  the sum over  wavevector components
such as $q_x$ runs over values $0, \pm 2\pi/L, \pm4\pi/L, \hdots$. The
Fourier components of $q_{\alpha\beta}({\bf q})$ at non-zero values of
${\bf q}$ are in this context \lq massive' modes and can be traced out
perturbatively.   Only the  ${\bf q}=0$  component has  to  be treated
non-perturbatively. Above  the upper critical  dimension the arguments
of  Br{\'e}zin  and Zinn-Justin  \cite{BZ}  show  that  the effect  of
integrating out  the non-zero ${\bf q}$  modes is simply  to shift the
mean-field  transition  temperature  to  the true  transition  and  to
renormalize the value of the  coupling constant $w$. The properties of
the ${\bf  q}=0$ fields, like those  in Binder ratio  plots studied in
simulations,  can then all  be extracted  from the  partition function
without the gradient terms
\begin{eqnarray}
Z&=&\int       \prod_{\alpha<\beta}      \left(\frac{dq_{\alpha\beta}}
{\sqrt{2\pi}}\right)\exp        \Big[        -\frac       {L^d\tau}{4}
\sum_{\alpha,\beta}q_{\alpha\beta}^2  \nonumber  \\  &&\quad\quad\quad
+\frac                                                        {L^dw}{6}
\sum_{\alpha,\beta,\gamma}q_{\alpha\beta}q_{\beta\gamma}q_{\gamma\alpha}
\Big].
\label{Zn}
\end{eqnarray}
A typical ratio would be
\begin{equation}
M_6=\frac{\langle (Tr q^3)^2 \rangle}{(\langle Tr q^2 \rangle)^3}.
\label{ratio}
\end{equation}
where the  thermal averaging denoted  by the angular brackets  is done
using  the partition function  of Eq.   (\ref{Zn}).  By  rescaling the
$q_{\alpha\beta}$ fields it is  easy to see that $M_6=g(N\tau^3/w^2)$.
The scaling functions  like $g(x)$ are the quantities  of interest and
can be calculated from  suitable derivatives of the partition function
$Z$.  The  neglect  of higher  terms  such  as  the quartic  terms  of
Eq. (\ref{FT}) is easily justified in the finite size critical scaling
region where  $N \rightarrow \infty$,  $\tau \rightarrow 0$,  but with
$N\tau^3/w^2$ finite.

 The partition  function $Z$ has recently been  extensively studied in
Ref.   \cite{YAM}. For earlier  studies see  Ref.  \cite{PRS}.  On the
high-temperature side of the transition i.e. when $\tau >0$, we showed
that  the  series  expansion  in  the  variable  $w^2/N\tau^3$,  while
formally a divergent  series, looks as though it  could be resummed to
give  useful results. The  technique adopted  to study  the high-order
terms in  the perturbation expansion was  to map the  problem onto the
spherical Ising  spin glass in the  Sherrington-Kirkpatrick (SK) limit
\cite{SK},  Our  concern  in  this   paper  is  what  happens  on  the
low-temperature side when $\tau<0$.

Let us  examine the first term  in Eq. (\ref{SE}) for  finite $N$.  In
sums over  ${\bf q}$, the ${\bf  q}=0$ terms coming  from the replicon
propagators are infinite unless we  keep $n$ finite since the replicon
propagator at  ${\bf q}=0$ is  just $1/nwQ$.  Thus the  most divergent
part in Eq. (\ref{SE}) is  of order $w^2/N(nwQ)^2$ and arises from the
term $I_{RR}$.  On  the other hand, the sums  over non-zero $q$ values
can  be   approximated  by   the  expressions  previously   given  for
$\Sigma_R(0)$.  Both  types of  terms make $\Sigma_R(0)$  positive and
apparently indicate  that expanding about the  replica symmetric state
would  lead  to  the   usual  instabilities.   However,  the  apparent
divergence of terms like $w^2/N(nwQ)^2$ as $n \rightarrow 0$ has first
to be resolved.

Such divergences  occur in  all orders of  the expansion and  the most
divergent terms come when all the propagators are of replicon type. At
$K$th order the  general form of these contributions  from ${\bf q}=0$
in the replicon sector is $(w^2/N(nwQ)^2)^K$ and retaining these terms
alone gives terms of the schematic form
\begin{eqnarray}
G_R(0)^{-1}&\sim   &   nwQ-\frac{w^2}{N(nwQ)^2}+\frac{w^4}{N^2(nwQ)^5}
+... \nonumber\\ &=&nwQf(w^2/N(nwQ)^3).
\label{RD}
\end{eqnarray}

Thus if the function $f(x)$ has the large $x$ form, $x^{\frac{1}{3}}$,
$G_R(0)\sim  N^{\frac{1}{3}}/w^{\frac{2}{3}}$ and the  problematic $n$
dependence  would disappear.  The  argument below  shows that  this is
indeed what happens. It arises  from the mapping to the spherical spin
glass in the SK limit \cite{YAM}. $G_R(0)$ physically is equal to
\begin{equation}
G_R(0)=\frac{1}{N}\sum_{i,j}\Big[\langle  S_iS_j   \rangle  -  \langle
 S_i\rangle \langle S_j \rangle \Big]^2.
\label{def}
\end{equation}
$G_R(0)$ can  be studied directly in  the spherical spin  glass in the
low-temperature regime. It can be approximated using the procedures in
\cite{YAM} by
\begin{equation}
G_R(0)=\frac{1}{N}{\sum_{\lambda}}^\prime \frac{1}{\lambda^2},
\label{SM}
\end{equation} 
where  the eigenvalues  $\lambda$ are  related  to those  of a  random
symmetric $J_{ij}$ matrix, such that in the large $N$ limit
\begin{equation}
\rho(\lambda)=\frac{N}{2\pi}[\lambda(4-\lambda)]^{\frac 12},
\label{rho}
\end{equation}
and  the prime in  the sum  indicates that  the smallest  (or negative
eigenvalues) are  to be omitted from  the sum in the  finite $N$ case.
Using the infinite $N$ form for the density of states the integral for
$G_R(0)$  would  appear to  be  divergent  from  the behavior  of  the
integrand at small  $\lambda$. However, the integral can  be cutoff at
the smallest eigenvalue,  which is of order $1/N^{2/3}$,  when one can
see that $G_R(0)$ is indeed  of order $N^{1/3}$. In principle by doing
numerical work on  the finite N spherical model,  it would be possible
to calculate $G_R(0)$ in the  finite size scaling limit to any desired
accuracy.

Returning  now to the  loop calculation,  the problematic  terms which
became infinite as $n$ goes to zero have been resummed so that now
\begin{equation}
G_R(0)^{-1}\sim w^{\frac 23}/N^{\frac 13}-\tau^2.
\label{res}
\end{equation}
The second  term of order $\tau^2$  comes from the  non-zero ${\bf q}$
contributions  in the  one-loop  calculation (  the  quartic terms  in
Eq. (\ref{FT})  also give terms  of order $\tau^2)$).  Notice  that in
the  finite   size  scaling  region  $N   \rightarrow  \infty$,  $\tau
\rightarrow 0$, but  with $N\tau^3$ fixed, the term  of order $\tau^2$
is  of order $1/N^{2/3}$  and so  is negligible  in comparison  to the
first term at large $N$.  The fact that it is negligible is what would
have been expected  anyway from the arguments of  Ref.  \cite{BZ}, who
showed that the only role  of the non-zero ${\bf q}$ contributions was
to  renormalise  the mean-field  transition  temperature and  coupling
constant  $w$. Notice that  the two  terms on  the right-hand  side of
Eq. (\ref{res}) become comparable when  $N\tau^6$ is $O(1)$. In the SK
limit, the number of metastable  (TAP) states $N_s$ goes like $\ln N_s
\sim N\tau^6$ for small  $\tau$ \cite{BM80} so that $G_R(0)^{-1}$ only
goes  negative  indicating  an  instability towards  replica  symmetry
breaking when  a multiplicity of states  exist - a  result entirely in
accord with the natural expectation.

Thus by going beyond perturbation  theory we have tamed divergences in
such  a  way  that  a  calculation which  seemingly  required  replica
symmetry  to be  broken to  cure an  instability present  at  one loop
order,  is no  longer unstable  when the  divergences are  summed. The
divergences in the finite size calculation, when resummed, give in the
large $N$ limit a negligible contribution if one is not working in the
finite  size  scaling  limit.   However,  it is  my  belief  that  the
divergences  which plague  the loop  expansion in  the low-temperature
state  for dimensions  $d<6$ as  $n$ goes  to zero  require  a similar
non-perturbative  treatment, and if  this could  be done,  the replica
symmetric state would emerge as stable.

Very  recently  numerical  evidence  \cite{KY}  has  emerged  that  is
consistent  with our  proposal that  replica symmetry  breaking occurs
only  when $d>6$.   The  one-dimensional long-range  Ising spin  glass
model    with    interactions     whose    magnitude    decrease    as
$1/r_{ij}^{\sigma}$  was studied.  When  $1/2 <  \sigma \le  2/3$, the
system is  expected to  behave like the  short-range spin  glass model
with $d>6$ and an AT line was  found in this interval.  No AT line was
found  when $2/3 <\sigma  \le 1$,  the interval  which is  expected to
mirror the short-range spin glass below six dimensions.  These results
are also consistent with earlier analytical calculations \cite{M86}.

Over  many years Newman  and Stein  have proved  a number  of rigorous
theorems  concerning  the  nature  of  the  ordered  phase  in  finite
dimenensional spin glasses \cite{NS}.  For $d<6$ our proposal that the
ordered  phase is  replica  symmetric and  droplet-like is  completely
consistent with  their theorems.  For  $d>6$ their \lq  chaotic pairs'
picture   fits  with   our  proposal   (but  there   are   also  other
possibilities).  In the chaotic  pair picture the domain wall exponent
$\theta>0$ but  its fractal  dimension $d_s =d$.   In other  words the
domain walls are space-filling. Also their theorems do not rule out an
AT  line  when $d_s=d$.   The  {\it  global}  Parisi overlap  function
$P(q)$,   which  is   closely   related  to   the  spatially   uniform
order-parameter  $q_{\alpha\beta}$,  could   be  non-trivial  for  the
chaotic pair  state, possibly even RSB-like.  They  emphasize in their
work though that the global $P(q)$ is not useful for understanding the
properties of pure states in finite dimensional glasses.

I should  like to acknowledge the  hospitality of the  Aspen Center of
Physics, Colorado,  USA during the writing  of this paper.   I am also
very grateful to  Dan Stein for patiently explaining  his results with
Newman to me.

\end{document}